\documentclass[a4paper,11pt]{article}
\usepackage{pos}
\usepackage{bbold}
\usepackage{subfig}
\usepackage{pgf}
\title{Determination of $m_c$ from $N_f = 2+1$ QCD with Wilson fermions}
\manuallySeparateAuthors
\author[a]{Gunnar Bali}
\author*[a,b]{, Sjoerd Bouma}
\author[a]{, Sara Collins}
\author[a]{and Wolfgang S\"oldner}
\author{\\ (RQCD Collaboration)\\[0.5cm]
  \hspace{3mm}\includegraphics[width=3cm]{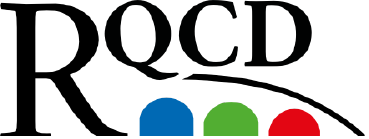}}
\renewcommand{\printHeadAuthors}{Sjoerd Bouma} 
\affiliation[a]{Institut f\"ur Theoretische Physik,\\
  Universtät Regensburg,
  93040 Regensburg, Germany}

\affiliation[b]{ECAP,\\ FAU Erlangen-N\"urnberg,
91058 Erlangen, Germany}

\emailAdd{sjoerd.bouma@fau.de}
\emailAdd{gunnar.bali@ur.de}
\emailAdd{sara.collins@ur.de}
\emailAdd{wolfgang.soeldner@ur.de}

\abstract{We present preliminary results for the charm
  quark mass in the $N_f=4$ RGI scheme. These were
  obtained using $N_f=2+1$ CLS ensembles with $\mathcal{O}(a)$
  non-perturbatively improved Wilson fermions. We employed
  five different lattice spacings, ranging down to $a\lesssim 0.04$~fm
  and realized
  approximately physical pion and kaon masses, with ensembles spread out
  along three different trajectories in the quark mass plane, enabling
  a thorough study of the dependence on the lattice spacing and the
  light and strange sea quark masses. We sketch our analysis
  strategy and find that the dominant errors at present
  are due to the renormalization and scale setting uncertainties.}

\FullConference{%
 The 38th International Symposium on Lattice Field Theory, LATTICE2021
  26th-30th July, 2021
  Zoom/Gather@Massachusetts Institute of Technology
}

\begin{document}
\maketitle
\section{Introduction}
\label{section:introduction}
The charm quark mass is of interest both as a fundamental parameter of
the Standard Model and as input to phenomenological predictions,
including for BSM physics. Here, we present preliminary results of a
recent analysis, determining the charm quark mass using the CLS $N_f =
2 + 1$ ensembles with $\mathcal{O}(a)$ improved Wilson-clover
fermions. Discretisation effects are normally significant for charm
observables in current simulations and the CLS ensembles, with the
squared lattice spacing varied by a factor of almost five, enable such
systematics to be tightly controlled.

\begin{figure}[ht]
    \centering
    \scalebox{.6}{\input{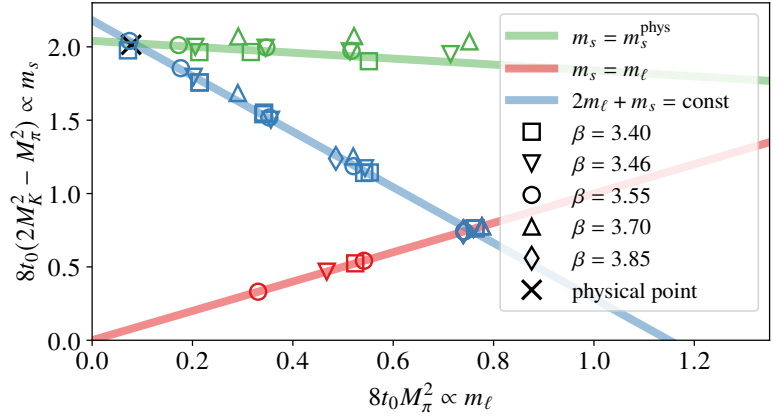}}
    \caption{Overview of the CLS ensembles used. The cross symbolizes physical pion and kaon masses.}
    \label{fig:phi2phi4}
\end{figure}
\section{Setup}
\label{section:setup}
We use 39 ensembles of non-perturbatively improved Wilson-clover fermions generated by the Coordinated Lattice Simulations (CLS~\cite{Bruno:2014jqa}) effort, using OpenQCD~\cite{Luscher:2012av}. These ensembles were generated at five different values of the lattice spacings, ranging from $a\approx 0.085$~fm down to $a\approx 0.039$~fm. Pion masses range from 420 MeV down to the physical point. The ensembles approximately lie on three different `chiral trajectories': 1) constant strange quark mass, where $m_s \approx m_s^\mathrm{phys}$; 2) symmetric line, where $m_s = m_\ell$; 3) constant sea quark mass, where the trace of the quark mass matrix $\mathrm{Tr}[M_q] = 2m_\ell + m_s =\mathrm{const.}$. The inclusion of multiple chiral trajectories allows us to compensate for any (slight) mistuning of individual trajectories in the fit. An overview of the sea quark mass combinations covered in this analysis is shown in figure~\ref{fig:phi2phi4}.
For each ensemble two heavy quark masses ($m_H$) around the physical charm quark mass were simulated. The charm quark mass ($m_c$) is then determined by an interpolation in the fit, as detailed in section~\ref{section:fitting}. Only ensembles with a spatial extent $L$ satisfying $L M_\pi \gtrsim 4$ were included in order to minimize finite-size effects.

\section{Analysis strategy}
\label{section:method}
The charm quark mass is determined using the PCAC relation:
\begin{align}
    am^\textrm{PCAC}(t) = \frac{\partial_t C_{A_0^L P^{(j)}}(t) + c_A a\partial_t^2 C_{P^L P^{(j)} }(t))}{2 C_{P^L P^{(j)}}(t)},
\end{align}
where $C_{A_0P}$ and $C_{PP}$ are point-smeared axial-pseudoscalar and pseudoscalar-pseudoscalar two-point correlation functions, the superscript $(j)$ represents the level of spatial smearing and the improvement coefficient $c_A$ was determined non-perturbatively in~\cite{ref:ca}. We use flavour non-diagonal currents  $J=\bar{q}_i\Gamma q_j$ with $\Gamma=\gamma_5 (\gamma_t\gamma_5)$ for $J=P(A)$  and $i,j\in\{u, d, s, H\}$ to obtain $m_{ij}^{\textrm{PCAC}}=\tfrac12(m_i^{\textrm{PCAC}}+m_j^{\textrm{PCAC}})$, where $m_u=m_d=m_{\ell}$ and $m_H\approx m_c$.
We employ two different definitions of the discrete derivative $\partial_t$: the `standard', symmetric discretized derivative $\partial_t f(t) = \tfrac{f(t+a) - f(t-a)}{2a}$ and a `continuum' definition of the PCAC mass based on the following parametrizations:
\begin{align}
    C_{PP}(t) &= \mathcal{Z}_{PP} \left( e^{-amt} + r e^{am (T-t)}\right) + \textrm{excited states},\\
    C_{AP}(t) &= \mathcal{Z}_{AP} \left( e^{-amt} - r e^{am (T-t)}\right) + \textrm{excited states},\\
    am^\mathrm{PCAC}(t) &= \frac{1}{2} am(t) \left( \frac{\mathcal{Z}_{AP}}{\mathcal{Z}_{PP}} + c_A am(t)\right),
\end{align}
where $r = 0$ for ensembles with open boundary conditions in time, and $r=1$ for periodic boundary conditions. Neglecting the excited states, we can then determine $m(t)$ and $\mathcal{Z}_{AP}/\mathcal{Z}_{PP}$:
\begin{align}
    am(t) = &= \begin{cases}
    \frac{1}{4} \left(\log \frac{C_{PP}(t-a)}{C_{PP}(t+a)} + \log \frac{C_{AP}(t-a)}{C_{AP}(t+a)}\right) \quad (r=0), \\
    \frac{1}{2} \left(\cosh^{-1} \frac{C_{PP}(t-a)+C_{PP}(t+a)}{2 C_{PP}(t)} + \cosh^{-1} \frac{C_{AP}(t-a)+C_{AP}(t+a)}{2 C_{AP}(t)}\right) \quad (r=1),
        \end{cases}\\
    \frac{\mathcal{Z}_{AP}}{\mathcal{Z}_{PP}}(t) &= \begin{cases}
    \frac{C_{AP}(t)}{C_{PP}(t)} \quad(r=0), \\
    \left(\frac{C_{AP}(t) \cdot (C_{AP}(t+a) - C_{AP}(t-a))}{C_{PP}(t) \cdot (C_{PP}(t+a)-C_{PP}(t-a) )}\right)^{1/2} \quad (r=1).
    \end{cases}
\end{align}

\begin{figure}
    \centering
    \scalebox{.6}{\includegraphics{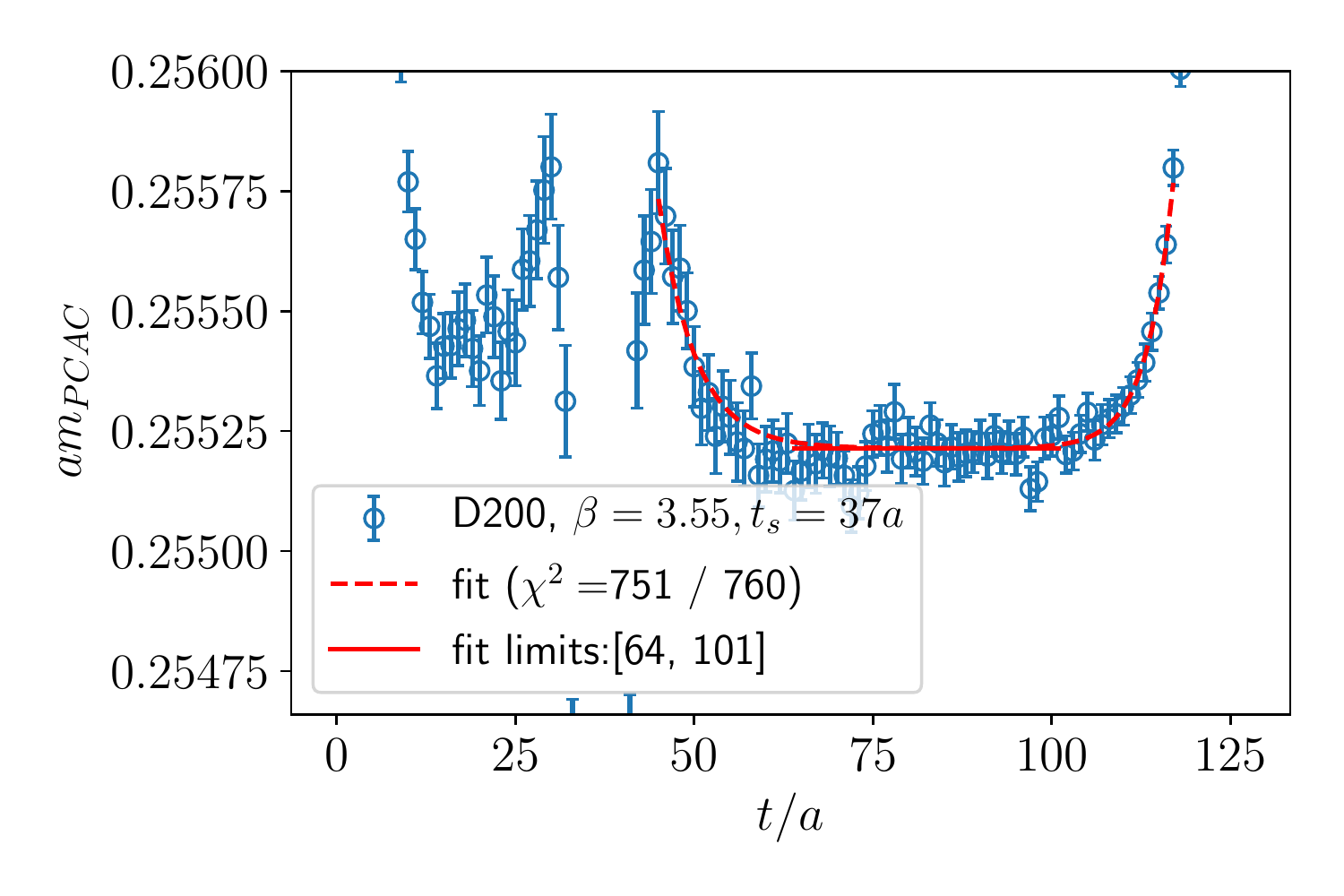}}
    \caption{Example of the fit that determines the plateau range. The fit is performed simultaneously to correlator data for multiple source positions, but only $t_s=37a$ is shown here.}
    \label{fig:D200_fit_range}
\end{figure}

To identify regions in $t$ where discretization and boundary effects for either definition of PCAC masses~---~symmetric derivative ($\partial_{\mathrm{std}}$) and continuum-inspired ($\partial_{\mathrm{con}}$)~---~can be neglected, the effective PCAC masses are fitted to a simple constant plus exponential form:
\begin{align}
    am^\mathrm{PCAC}(t) \approx am^\mathrm{PCAC} + c_1 e^{-b_1t} + c_2 e^{-b_2(T-t)},
    \label{eq:pcacfit}
\end{align}
where for periodic boundary conditions $b_2=b_1$ and $c_2=c_1$.
The plateau is then defined as the region where corrections to the constant are  smaller than a quarter of the statistical error, i.e.
$4 \cdot \left|c_1 \cdot \exp^{-b_1t}+ c_2 e^{-b_2(T-t)}\right| \leq \Delta^\mathrm{stat}[am^\mathrm{PCAC}(t)]$. 
For ensembles where multiple source positions are available, a simultaneous fit is performed for all source positions sufficiently far away from the boundary. An example of the results of one of these fits is shown in figure~\ref{fig:D200_fit_range}.
Once the plateau region is identified, the PCAC mass is computed as a simple weighted average, i.e.\ carrying out a one-parameter fit. Errors are estimated through a binned jackknife procedure in order to properly take into account autocorrelations: the jackknife error is computed after binning in Monte Carlo time for a number of different bin sizes $S$. The integrated autocorrelation time $\tau_\mathrm{int}$ is then estimated by extrapolating to infinite bin size:
\begin{align}
     \frac{\sigma^2[S]}{\sigma^2[1]} \approx 2 \tau_\mathrm{int} \left( 1 - \frac{c_A}{S} + \frac{d_A}{S}e^{-S/\tau_\mathrm{int}} \right).
\end{align}
The autocorrelation-corrected error is obtained by rescaling the error at bin size 1 by $\sqrt{2\tau_\mathrm{int}}$. A similar procedure was followed in order to obtain covariances between different observables used in the fits.

\section{Renormalization Group Invariant (RGI) masses}
\label{section:fitting}
The Renormalization Group Invariant (RGI) quark masses are obtained from the PCAC masses using the following relation:
\begin{align}
    m_{ij}^\mathrm{RGI} = 
            Z_M m_{ij}^{\mathrm{PCAC}} \left[ 1 + (b_A - b_P) a m_{q,ij} + (\tilde{b}_A - \tilde{b}_P) a \mathrm{Tr}[M_q] \right] 
            +  \mathcal{O}(a^2). \label{eq:RGI}
\end{align}
The values for the renormalization and improvement coefficients $Z_M$ and $b_A-b_P$ were determined non-perturbatively in~\cite{ref:ZM} and~\cite{ref:babp}, respectively. For the value of $\tilde{b}_A - \tilde{b}_P$, no precise non-perturbative determination is available, but as the term it multiplies is proportional to the sea quark masses in lattice units, which are tiny compared to the heavy mass $am_{q, HH}=\tfrac12(\tfrac{1}{\kappa_H}-\tfrac{1}{\kappa_{\mathrm{crit}}})$, we ignore this term and set $\tilde{b}_A - \tilde{b}_P = 0$ in our fits. The heavy quark mass can be obtained either from $m_{HH}$, from $m_{Hs}$ or from $m_{H\ell}$: $m_H=m_{HH}$, $m_H=2m_{Hs}-2m_{s\ell}+2m_{\ell\ell}$ or $m_{H}=2m_{H\ell}-2m_{\ell\ell}$, where $m_{\ell\ell}=m_{ud}$ is a flavour non-singlet combination and so is $m_{HH}$ because the heavy quark is quenched.

As the ensembles were generated at fixed values of the bare coupling $g_0$, rather than of the order-$a$ improved coupling, in order to maintain $\mathcal{O}(a)$ improvement, all masses are rescaled by the Wilson flow scale $t_0$, for which we use the notation $\mathbb{m} = \sqrt{8t_0}m$. The continuum and chiral extrapolation is performed through a global, fully correlated fit. As the statistical errors of the heavy PCAC masses are much smaller than those of, e.g., the pion and kaon masses, we use a generalized chi-squared fit, in which the errors of $M_\pi$ and $M_K$ are included, as well as their correlations with the $D$ meson mass and the heavy quark mass. Errors and covariance matrix are estimated through the procedure outlined in section~\ref{section:method}. We include a dependence on either $m_{D_s}$ or the flavour-averaged $D$-meson mass $m_{\overline{D}} = \tfrac{2m_D + m_{D_s}}{3}$, in order to allow for a global interpolation from the two simulated heavy quark masses to the physical charm quark mass. The values of the improvement coefficients $Z_M, b_A-b_P$ as well as of the scale $t_0/a$ are added as priors to the $\chi^2$ functional, along with their uncertainties.
The full fit parametrization is given below:
\begin{align}
    m_{H}^\mathrm{RGI}(a^2/t_0^*, \overline{\mathbb{M}}^2, \delta\mathbb{M}^2, { \mathbb{m}_D}) &= 
    \begin{cases} f_\mathrm{\chi PT} \times (1 + f_\mathrm{latt}), \\
    f_\mathrm{\chi PT} + f_\mathrm{latt},
    \end{cases}, \qquad \textrm{where}\\ 
    f_\mathrm{\chi PT}(\overline{\mathbb{M}}^2, \delta\mathbb{M}^2, \mathbb{m}_D) &= p_0  + p_1 \overline{\mathbb{M}}^2 + p_2 \delta\mathbb{M}^2 + p_7 { \delta\mathbb{m}_D}
    + p_3 \overline{\mathbb{M}}^4 + p_4 \delta\mathbb{M}^4 \\ \notag
    &\phantom{=} + p_8 { \mathbb{m}_D} \overline{\mathbb{M}}^2 + p_9 { \mathbb{m}_D} \delta\mathbb{M}^2 + p_{10} { \delta\mathbb{m}_D^2} + p_{11} \overline{\mathbb{M}}^2 \delta\mathbb{M}^2 + p_{14} { \delta\mathbb{m}_D^3},\\
    f_\mathrm{latt}(a^2/t_0^*, \overline{\mathbb{M}}^2, \delta\mathbb{M}^2, { \mathbb{m}_D}) &= \frac{a^2}{t_0^*} \Big[ p_{15} + p_{16} \overline{\mathbb{M}}^2 + p_{17} \delta\mathbb{M}^2  + p_{20} {\delta\mathbb{m}_D }\\ \notag
    &\phantom{=\Big[\frac{a^2}{t_0^*}}+  p_{24} {\delta\mathbb{m}_D^2} + p_{27} {\mathbb{m}_D} \overline{\mathbb{M}}^2 + p_{28} { \mathbb{m}_D} \delta\mathbb{M}^2 \Big]  \\ \notag
   &\phantom{=} + \left(\frac{a^2}{t_0^*}\right)^k \left[ p_{18} + p_{21} {\delta\mathbb{m}_D} + p_{25} \overline{\mathbb{M}}^2  + p_{26} \delta\mathbb{M}^2 \right].
\end{align}
Here, $\overline{\mathbb{M}}^2 = \frac{2\mathbb{m}_K^2 + \mathbb{m}_\pi^2}{3}$, $\delta \mathbb{M}^2 = 2 (\mathbb{m}_K^2 - \mathbb{m}_\pi^2)$, $\delta\mathbb{m}_D = \mathbb{m}_D - \mathbb{m}_D^\mathrm{phys}$, and $\mathbb{m}_D$ corresponding either to the $D_s$ or the $\overline{D}$. The lattice spacing dependence is parametrized through $\frac{a^2}{t_0^*}$, where $t_0^*$ is defined as the value of $t_0$ along the symmetric $m_s=m_\ell$ line which satisfies $8t_0^* (M_K^2 + M_\pi^2/2) = 12 t_0^*M_\pi^2 = 1.110$. This has the advantage over $t_0$ that it does not depend on the sea quark masses. Depending on the input, some parameters are set to zero. For instance, $p_2=0$ if $m_{\overline{D}}$ is used as an input since this term is only possible in conjunction with $m_{D_s}$. 
The physical charm quark mass $m_c^\mathrm{RGI}$ is then determined by setting $a^2/t_0^* = 0$ and the pion, kaon and $\overline{D}|D_s$ meson masses to their physical values, with $t_0^\mathrm{phys}$ taken from~\cite{Bruno:2016plf}.

The fit parametrization was varied by excluding different subsets of parameters. In order to arrive at a final result for each combination of PCAC flavour ($HH$, $Hs$ or $H\ell$), $D$-meson ($D_s$ or $\overline{D}$) and derivative ($\partial_{\mathrm{std}}$ or $\partial_{\mathrm{con}}$), we carry out a weighted average of the results over the parametrizations used, based on the Akaike Information Criterion (AIC) with the weight of a fit $k$ given by:
\begin{align}
  w_k = \frac{\exp [-\chi^2(k)/2-N_p(k)]}{\sum_n \exp [-\chi^2(n)/2-N_p(n)]},
\end{align}
where $N_p(k)$ is the number of parameters used. This procedure, which was also used, e.g., in~\cite{Heitger:2021apz}, allows us to investigate the systematic error associated to the parametrization. In total, roughly $\sim 100$ fits were carried out for each combination of PCAC flavour/$D$-meson/derivative.

\section{Results and discussion}
\label{section:results}
\begin{figure}
    \centering
    \scalebox{.6}{\input{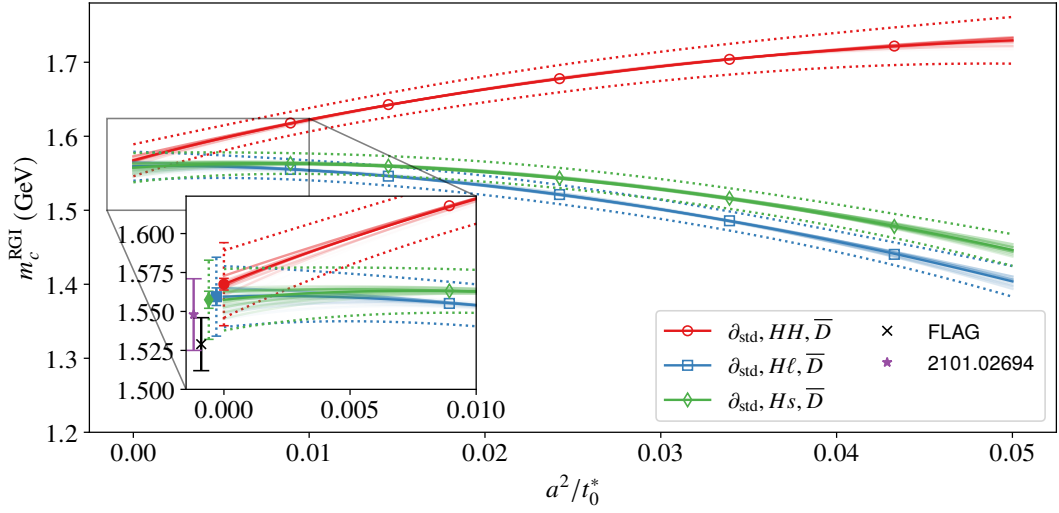}}
    \caption{Lattice spacing dependence for the fits with the standard discretized derivative, with the physical charm quark mass determined by interpolation to $\mathbb{m}_{\overline{D}} = \mathbb{m}_{\overline{D}}^\mathrm{phys}$. Different curves of the same colour correspond to different fit parametrizations, with their opacity proportional to their relative AIC weights. Solid error bars show the error due to the variation of the fit parametrization only, dotted curves and error bars show the total error.}
    \label{fig:AIC_lattice_spacing}
\end{figure}
\begin{figure}
    \centering
    \scalebox{.6}{\input{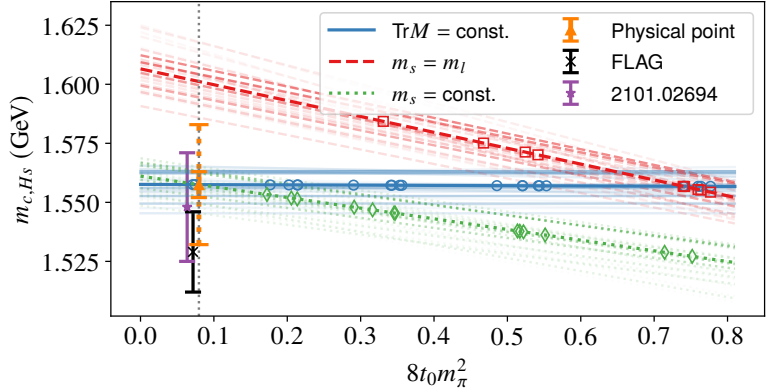}}
    \caption{Chiral dependence for the fits with the standard derivative, PCAC current $Hs$ and physical point determined by $\mathbb{m}_{\overline{D}}$. Two of the quark mass trajectories intersect at the physical point.}
    \label{fig:AIC_chiral_dependence}
\end{figure}
\begin{figure}
    \centering
    \scalebox{.6}{\input{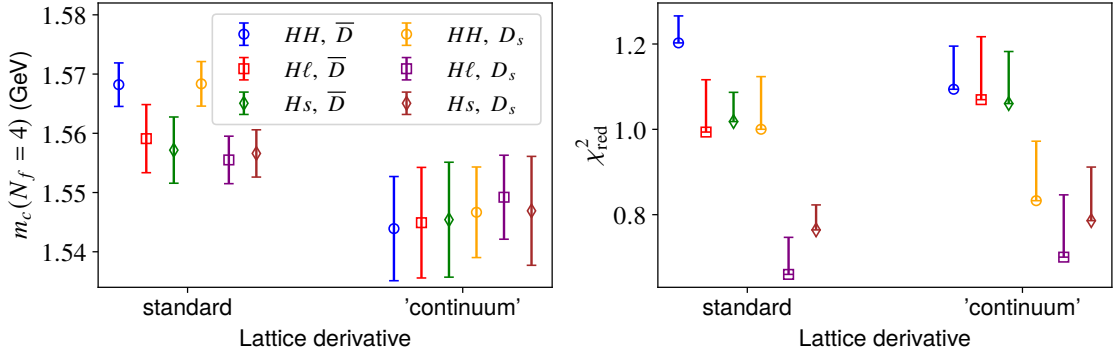}}
    \caption{(Left) Preliminary results with systematic errors due to fit variation only, for each derivative, flavour combination and $D$-meson. (Right) The range of $\chi^2_\mathrm{red}$ values for the fits. Only fits with a weight $w>0.01$ are shown on this panel.}
    \label{fig:AIC_overview}
\end{figure}
Preliminary results of our analysis are shown in figures~\ref{fig:AIC_lattice_spacing}--\ref{fig:AIC_overview}. Figure~\ref{fig:AIC_lattice_spacing} illustrates the dependence on the lattice spacing for the three different PCAC flavour combinations. The curves correspond to different fit parametrizations, with the opacity proportional to the weight in the AIC procedure described in section~\ref{section:fitting}. The overall errors are shown as dotted lines. For comparison, we have converted our results to the 4-flavour scheme, and also show the corresponding values from the FLAG 2019 report~\cite{FlavourLatticeAveragingGroup:2019iem} and the M\"unster 2020 result~\cite{Heitger:2021apz}. The latter is based on a subset of the CLS ensembles used here. The chiral extrapolation is shown in figure~\ref{fig:AIC_chiral_dependence}. For all the parametrizations explored, the dependence on the sea quark masses is much smaller than the lattice spacing effects.

An overview of all our results, only including the systematic error due to the parametrization, is shown in figure~\ref{fig:AIC_overview}. Agreement within these errors is moderate; there seems to be a systematic shift of results between the two different derivatives and in particular the values for the standard derivative in conjunction with the $HH$ PCAC current are somewhat larger than the other results. This may be related to the fact that discretisation effects for this combination are significantly larger than for the other combination, due to the large meson mass. The $\chi^2_\mathrm{red}$ values for the determinations using $\mathbb{m}_{\overline{D}}$ to set the physical charm quark mass are generally around 1; those using the $D_s$ mass instead are occasionally significantly lower, perhaps indicating some degree of overfitting. We remind the reader that in figure~\ref{fig:AIC_overview} only the errors due to the choice of parametrization are shown. It turns out that the overall error is dominated by the uncertainties of the renormalization ($Z_M$), the scale setting ($t_0^\mathrm{phys}$) and~---~to a lesser extent~---~the order-$a$ improvement coefficient $b_A-b_P$. The error budget, given for the determination using $Hs, \overline{D}, \partial_\mathrm{std}$, is shown in table~\ref{tab:errorbudget}. In relation to the overall error, the variation of the results between the twelve different determinations is small.\\

 \begin{table}
        \centering
        \begin{tabular}{l|r}
        \textbf{Error budget} (contribution to $\sigma_\mathrm{tot}^2$)& \\
        \hline
             Statistical (PCAC mass, $M_\pi, M_K, m_D, t_0$) &  9\%\\
             $\mathcal{O}(a)$ improvement ($b_A - b_P$) &  19\%\\
             Renormalization ($Z_M$) & 11\%\\
             Scale setting ($t_0^\mathrm{phys}$) & 21\%\\
             Renormalization scale & 35\%\\
             $N_f=3 \rightarrow 4$ conversion & 1\%\\
             \hline
             Fit parametrization & 5\%\\
        \end{tabular}
        \caption{Error budget for the RGI mass determination using
          $Hs, \overline{D}, \partial_\mathrm{std}$. Note that all but three sources of error (PCAC mass, $m_D$, fit parametrization) are shared between the different determinations
          of $m_c^{\mathrm{RGI}}$.}
        \label{tab:errorbudget}
    \end{table}

{\bf\noindent Acknowledgements.}
We thank Simon Kuberski, Jochen Heitger and Fabian Joswig for helpful
discussions and our CLS colleagues for the joint generation
of the gauge ensembles.
The authors were supported by the European Union’s Horizon 2020
research and innovation programme under the Marie Skłodowska-Curie
grant agreement nos.\ 813942 (ITN EuroPLEx) and
824093 (STRONG-2020) and by the Deutsche Forschungsgemeinschaft
(SFB/TRR-55).  The ensembles were generated as part of the CLS effort
using OpenQCD~\cite{Luscher:2012av}, and further analysis was performed
using a modified version of CHROMA~\cite{Edwards:2004sx}, the IDFLS
solver of OpenQCD and a multigrid
solver~\cite{frommer2014adaptive}. The authors gratefully acknowledge
the Gauss Centre for Supercomputing (GCS) for providing computing time
through the John von Neumann Institute for Computing (NIC) on
JUWELS~\cite{juwels} and on JURECA-Booster~\cite{jureca}
at Jülich Supercomputing
Centre (JSC). Part of the analysis was performed on the
QPACE~3 system of SFB/TRR-55 and the Athene cluster of the
University of Regensburg.
\bibliographystyle{JHEP}
\setlength{\bibsep}{0pt plus 0.3ex}
\bibliography{references.bib}
\end{document}